# Hiden Topics in Robotic Process Automation -- an Approach based on AI

Petr Průcha[1] [0000-0003-2197-7825]
Technical University of Liberec, Liberec, Czechia
petr.prucha@tul.cz

Peter Madzík [0000-0002-1655-6500]
Technical University of Liberec, Liberec, Czechia
peter.madzik@gmail.com

Lukáš Falát [0000-0002-2597-7059]
University of Žilina, Žilina, Slovakia
lukas.falat@fri.uniza.sk

**Abstract**
Robotic Process Automation (RPA) has rapidly evolved into a widely recognized and influential software technology. Its growing relevance has sparked diverse research efforts across various disciplines. This study aims to map the scientific landscape of RPA by identifying key thematic areas, tracking their development over time, and assessing their academic impact. To achieve this, we apply an unsupervised machine learning technique—Latent Dirichlet Allocation (LDA)—to analyze the abstracts of over 2,000 scholarly articles. Our analysis reveals 100 distinct research topics, with 15 of the most prominent themes featured in a science map designed to support future exploration and understanding of RPA's expanding research frontier.

**Keywords:** Robotic Process Automation, RPA, Hyper-Automation, Intelligent Automation, Smart Literature Review, LDA

# 1 Introduction

Enhancing business competitiveness increasingly depends on continuous, incremental improvements. Organizations striving for greater efficiency, cost and time savings, and more transparent decision-making processes often turn to automation tools—particularly those grounded in Robotic Process Automation (RPA). RPA is a software-based technology that employs software robots (or "bots") to automate repetitive, rule-based business tasks (Ivančić et al., 2019; Willcocks et al., 2015). These bots are capable of replicating human interactions with digital systems, executing predefined processes autonomously without human intervention.

The appeal of RPA lies in its ability to mimic human actions, its rapid deployment, ease of use, and its capacity to automate up to 95% of user activities. These advantages position

[1] Corresponding author

RPA as a more accessible and scalable solution compared to traditional automation approaches (Czarnecki & Fettke, 2021; Lhuer, 2016; Syed et al., 2020; van der Aalst et al., 2018).

In recent years, RPA has seen rapid adoption, particularly in sectors such as finance and telecommunications (Lacity et al., 2016; Lacity & Willcocks, 2016). Its use has since expanded across a broad range of industries, with administrative functions being a common area of implementation (Wewerka & Reichert, 2021). However, RPA's reliance on software environments—most notably Windows—limits its applicability to processes that can be executed within that ecosystem, narrowing its scope to certain market segments.

Despite this constraint, RPA continues to evolve. Beyond business process automation, it is now being employed in areas such as software testing and healthcare. In the latter, RPA bots facilitate data transfer between departments, helping to streamline operations and improve service delivery (Hewitt et al., 2021; Holmberg & Dobslaw, 2022).

Even though the growth rate of the RPA market in the last 5 years is slowing down (Fersht, 2018; Virbahu Nandishwar Jain & DAZ Systems, 2019), in the year 2030, the size of the market will be considerable: it is predicted to be around 20 billion USD (Spherical Insights, 2023; Vailshery, 2022). RPA bots can be extended by extensions from other companies or the RPA community for example the UiPath marketplace has over 1700 extensions (UiPath Marketplace, 2023.). The extension can be used for various reasons, for example OCR readers, converting and checking PDFs, logging and monitoring activities, preparing connectors to apps (CRMs, ERPs, office suites) and many others. The flexibility to extend by additional applications gives RPA the power to leverage the current trends as artificial intelligence (AI) with large language models (LLMs). With intelligent automation and hyper-automation trends, RPA has the potential to grow and serve as a tool for automation and as an agent to connect multiple applications (Gartner Inc., 2021).

RPA technology is widely adopted across many fields and used in various applications. For future development of robotic process automation is crucial to identify the current trends and form the future direction of robotic process automation.

We analyse recent systematic literature reviews and bibliographic reviews on the topic of RPA in Table 1. All papers use a manual analysis of full texts and a selection of papers. Choosing papers manually through selection rules in the document can be arbitrary and lack determinism. This significantly restricts literature reviews, preventing the inclusion of extensive amounts of studies for analysis. The studies in Table 1 do not analyse the trend of RPA progressing over time. Which makes it impossible to find hidden past and future trends in the field of RPA. Analysing the data manually makes it challenging to discern the broader perspective and essential interconnections among the domains.

Table 1. Analysis of systematic literature reviews and bibliographic reviews

| Author | Number of Papers | Results |
| --- | --- | --- |

| Stravinskienė & Serafinas, 2021 | 29 papers | Presented insight to focus of RPA studies, to the primary sector of RPA and region of RPA use. |
|---|---|---|
| Wewerka & Reichert, 2021 | 63 papers | Create 4 thematic clusters of RPA studies. |
| Syed et al., 2020 | 125 papers | Propose 15 future research challenges in the RPA domain. |
| Enriquez et al., 2020 | 54 papers | Growing research interest in RPA and analysis of RPA platforms. |
| Ivančić et al., 2019 | 36 papers | Presented insight into the benefits of RPA. |

The number of research papers related to RPA is growing dramatically fast, and development is progressing at a very dynamic pace. With the current scope, it is no longer practically possible to manually make a complex science map of RPA research. Current review studies related to RPA research work with a limited number of papers, which may distort the results. The dynamics of RPA-based research development have resulted in the gradual fragmentation of research into sub-areas, the development and intensity of which have so far remained unrevealed. A science map related to RPA research can help capture the current and future aspects of RPA by systematising latent topics and thus open a discussion about research directions and their potential. So far, the systematisation of latent (or even explicit) topics has been carried out through SLR, which limited the analysis to a few dozen to a hundred studies, which were examined manually. The potential of in-depth analysis of topics from a much larger number of research papers was not used until recently. However, current machine learning algorithms offer a fast and relatively reliable opportunity to process unstructured data in a way that allows for deep analysis (Asmussen & Møller, 2019).

The aim of this study is to create a comprehensive science map of RPA by revealing latent topics related to RPA, their research interest, impact and time development. We want to achieve such a science map with a set of analyses, which can be characterised by three research questions (RQ):

RQ1: What are the trends and structure of research related to RPA?

RQ3: What are the most significant topics, and how is their evolution over time?

In our study, we analysed more than 2,000 abstracts of research papers through an innovative approach based on machine learning algorithms. The basis of the analysis was the processing of unstructured data (i.e. text) by systematic application of Latent Dirichlet Allocation (LDA). By using LDA, it was possible to identify unique research topics, their research interest, and their impact on a large number of documents, as well as to assess their development over time. Together, the analyses in this study form a current RPA research state-of-the-art and can contribute to a broader discussion regarding the future direction of RPA, its limits, and application areas.

# 2 Methodology

To find out the current state of the scientific discipline of RPA, we used a smart literature review (Asmussen & Møller, 2019; Falát et al., 2023). Several dozen studies are typically used in systematic literature reviews. However, by using the smart literature review based on the machine learning approach, we were able to examine 2,351 studies found in the Scopus database in the preceding section because of the machine learning approach. The results' great dependability and representativeness—another benefit—follow from the preceding. The analysis consisted of three phases, a brief description of which can be found in the following subsections.

## 2.1 Data collection

Relevant research papers were exported from the Scopus database, which is one of the most extensive scientific databases and contains all necessary bibliometric information, including document abstracts. Data collection took place on August 11, 2023. To search for research papers related to RPA, the search string: "Robotic Process Automation" OR "Intelligent Process Automation" OR ( "Software Robots" AND "process automation" ) was used. Using this query, we identified 2,351 documents. After removing documents that did not have an abstract, the final dataset contained 2,299 documents. This dataset was merged with information about the subject areas of individual journals. With this merger, it was achieved that each document was assigned to one or more subject areas, depending on the source (journal, book, proceedings, etc.) in which it was published.

## 2.2 Descriptive analysis

Standard bibliometric approaches were used to answer RQ1. Trends in RPA research was monitored at annual intervals and research interest (nr. of published papers) and research impact (nr. of citations) were evaluated. The description also contained the identification of the most important subject areas related to RPA research and their research impact.

## 2.3 Latent topics extraction

To answer RQ2, LDA topic modelling was used, which belongs to the unsupervised machine-learning method for analysing textual data. The result of LDA is the identification of thematic clusters (topics) with a similar word structure. The LDA procedure itself contains three basic steps - sections 2.3.1 to 2.3.3.

### 2.3.1 Data pre-processing

At first, text pre-processing, which ensured more accurate results, was performed. A conventional text pre-processing process and specific text pre-processing were both part of the pre-processing step. After replacing some special characters with spaces as part of the regular pre-processing routine, we excluded punctuation, eliminated numerals, eliminated English stopwords defined in the tm package, and eliminated additional common English stopwords that were not included in the tm package in the R language. The words were then trimmed and extra spaces were removed. Afterwards, the text was subjected to tailored pre-processing, which included the elimination of terms (in appendix A) that did not have a specific meaning that was pertinent to the topic of RPA. These words were identified using frequency analysis. After stopwords elimination and stemming, the calculation of an optimal number of topics was carried out.

### 2.3.2 Finding the optimal number of topics

Finding the optimal number of topics was the first step in the LDA modelling process. We used a statistical approach based on four criteria to determine the ideal number of topics in the collection of abstracts (Arun et al., 2010; Cao et al., 2009; Deveaud et al., 2014; Griffiths & Steyvers, 2004). The maximum value of the sum function from the aforementioned metrics was used to calculate the optimal number of topics. Using the Gibbs sampling, the LDA method's parameters were quantified for every tested number of topics. The algorithm's number of iterations was set at 2,000 for each run. We only included every 200th iteration from the quantification to increase the robustness of our solution. We conducted five runs for each tested number of topics, *k*, taking into account just the top result for each *k*. In order to make our solution replicable, we established the following seed list. {7413, 32, 23935, 8461, 279}. In total, we tested 30 different numbers of topics - a sequence from 10 to 300 by a step increment of 10. Figure 2 shows the outcomes of the optimization procedure.

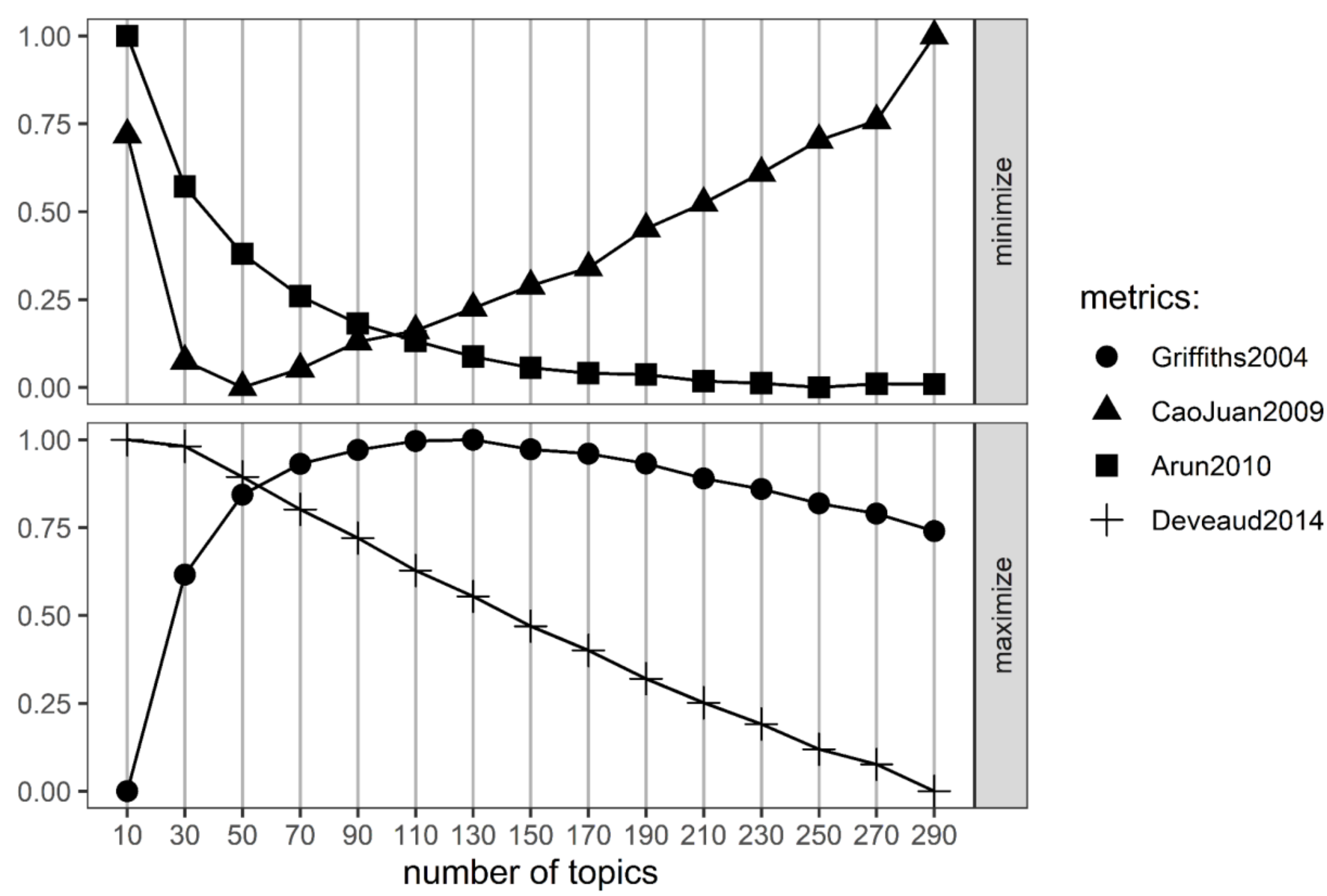

**Figure 2** Selection of optimal number of topics

The total function of the four tracked metrics was used to find the ideal number of topics, which was found to be 100, as shown in Figure 2. In consideration of the established statistical criteria, this suggests that the full set of research abstracts may be separated into 100 somewhat separate study topics (themes). Additionally, it is important to note that the probability-based topicalization of documents into groups of documents provides the foundation for choosing the ideal number of subjects to use LDA.

### 2.3.3 Topic modelling

Latent Dirichlet Allocation (LDA) is an unsupervised machine-learning method for analysing textual data. LDA is based on several principles. Firstly, LDA is a generative model based on probabilistic clustering. LDA is a hidden variable model. The model decomposes the text corpus into latent topics. The observed data are represented by the specific words in each document and hidden variables represent latent topics (Blei and Lafferty, 2009).

Secondly, it's important to note that in Latent Dirichlet Allocation (LDA), documents are not confined to a single topic. That is why LDA is called a mixed-membership model. Each document is a random mix of several topics (Blei et al., 2003). Documents belong to several topics at the same time (Grün and Hornik, 2011). Each topic is defined by a probability distribution over words, while it is assumed that the topics are uncorrelated (Grün & Hornik, 2011).

Thirdly, LDA is a "bag of words" model, i.e., the words in documents are interchangeable, and word order does not play a role in parsing (Blei & Lafferty, 2007).

Finally, part of its name comes from the Dirichlet distribution. The probability distribution of analysed documents (abstracts in our case) over topics as well as the probability distribution of words in individual documents, uses the Dirichlet probability distribution (Ponweiser, 2012). It is an exponential family distribution which has a density defined as

$$p(x|\vec{\alpha}) = \frac{\Gamma(\sum_{i=1}^{k} \alpha_i)}{\prod_{i=1}^{K} \Gamma(\alpha_i)} \prod_{i=1}^{K} \theta_i^{\alpha_i - 1} \quad (1)$$

where $\Gamma$ is a Gamma function and $\vec{\alpha}$ is a positive K-vector, where K is a specified number of topics.

More formally, the LDA model can be defined in the following steps (Blei et al., 2003; Blei & Lafferty, 2009; Ponweiser, 2012):

1. For every topic *k*, $k = 1 \ldots K$,
   a. determine the most probable words of the topic by defining a probability distribution of terms (words) per topic $\phi_k$. Let the variable $\phi_k$ have a Dirichlet distribution $Dir(\beta)$, where $\beta$ is a positive *V*-vector.
2. For every document *d*, (represented by a paper abstract in our case)
   a. determine what proportion of topics are likely to be in the document. Do this by defining a vector of topic proportions in a document denoted as $\theta_d$. Let the variable $\theta_d$ have a *K*-dimensional Dirichlet distribution $Dir(\alpha)$ where *K* is the number of words in a vocabulary.
   b. For each word $w_i$ in a document *d*

i. Choose an appropriate topic. Do this by defining a topic assignment $z_{d,i} \sim Multinomial(\theta_d)$. Note that $z_{d,i} \in \{1 \dots K\}$

ii. based on a selected topic, determine what word is likely. Choose a word $w_{d,i}$ from a multinomial probability conditioned on the topic $\phi_{z_{d,i}}$

Coming out from Blei et al. (2003), Ponweiser (2012) defines the joint distribution as the composition of topic mixture $\theta$, a set of topics denoted as *z*, terms distributions per topic $\phi$ and a set of words denoted as *w*. It is formally defined as

$$p(\alpha, \beta) = p(\beta)p(\alpha)p(\theta)p(w|z, \phi) \quad (2)$$

where $\alpha$ and $\beta$ are the hyperparameters of the model.

According to Ponweiser (2012) the ***probability distributions of terms per topic*** $\phi_{k,v}$, which is based on the Dirichlet distribution with $\beta$ parameter, define the probability that term *v* is drawn when the topic was chosen to be *k*. Let us then define the probability $\phi$ for all topics and all words of the vocabulary.

$$p(\beta) = \prod_k \frac{\Gamma(\beta_k)}{\prod_v \Gamma(\beta_{k,v})} \prod_v \phi_{k,v}^{\beta_{k,v}-1} \quad (3)$$

The ***probability distribution of topics over a document*** $\theta$, which is drawn from the Dirichlet distribution with $\alpha$ parameters, is defined according to Blei et al. (2003) and Ponweiser (2012) as

$$p(\vec{\alpha}) = \frac{\Gamma(\sum_k \alpha_k)}{\prod_k \Gamma(\alpha_k)} \prod_k \theta_k^{\alpha_{k-1}} \quad (4)$$

The ***topic to words assignment*** $z_{d,i}$ defines the affiliation of the $i^{th}$ word in the $d^{th}$ document to the specific topic. This affiliation depends on the distribution $\theta$, i.e., on topic proportions distribution for a specific document.

Let us define $n_{d,k}$ as how many times a given topic *k* is assigned to any word in document *d*. Then the probability of *z* for all topics and all documents is defined (Ponweiser, 2012) as follows.

$$p(\theta) = \prod_{d=1}^{D} \prod_{k=1}^{K} \theta_{d,k}^{n_{d,k}} \quad (5)$$

and the probability of a corpus *w* as (Ponweiser, 2012):

$$p(z, \phi) = \prod_{k=1}^{K} \prod_{v=1}^{V} \phi_{k,v}^{n_{k,v}} \quad (6)$$

## 2.4 Identification of the most significant topics

A combination of bibliometric tools and topic modelling results through LDA was used to answer RQ3. Each topic contained aggregated information on research interest and research impact. This information was plotted in a scatter diagram, while statistical boundaries were set for topic categorization - median, 3rd quartile, 90th percentile - for research interest and research impact. This made it possible to create groups of topics and, at the same time, to reveal those that are among the most significant in the field of RPA.

# 3 Results

## 3.1 Overview of papers related to RPA

The RQ1 was aimed at obtaining a basic overview of research in the field of robotic process automation. Figure 3 shows the development of the number of published papers compared to citations in individual years. From the long-term trend, a significant increase in research interest (measured through the number of published papers) can be seen from 2018 while this interest has been very significant in the last three years. The analysis of the trends goes up to 11 August of 2023, so there are no included results for the year 2023. For now, we can see a lower volume of publications than in 2022. Nevertheless, the number of publications is in August close to the year 2021. We suppose that the number of publications for 2023 will still grow.

If we compare the trend of research interest with the research impact trend (measured through the number of citations of papers), we can see that the surge is visible similarly to the number of published papers. The peak of citations is in the year 2021. In the years 2022 and 2023, we recorded a decrease in research impact, which will be very likely caused by the time lag of citations (Min et al., 2016). From both points of view, it can be concluded that robotic process automation is relevant and current for the scientific community.

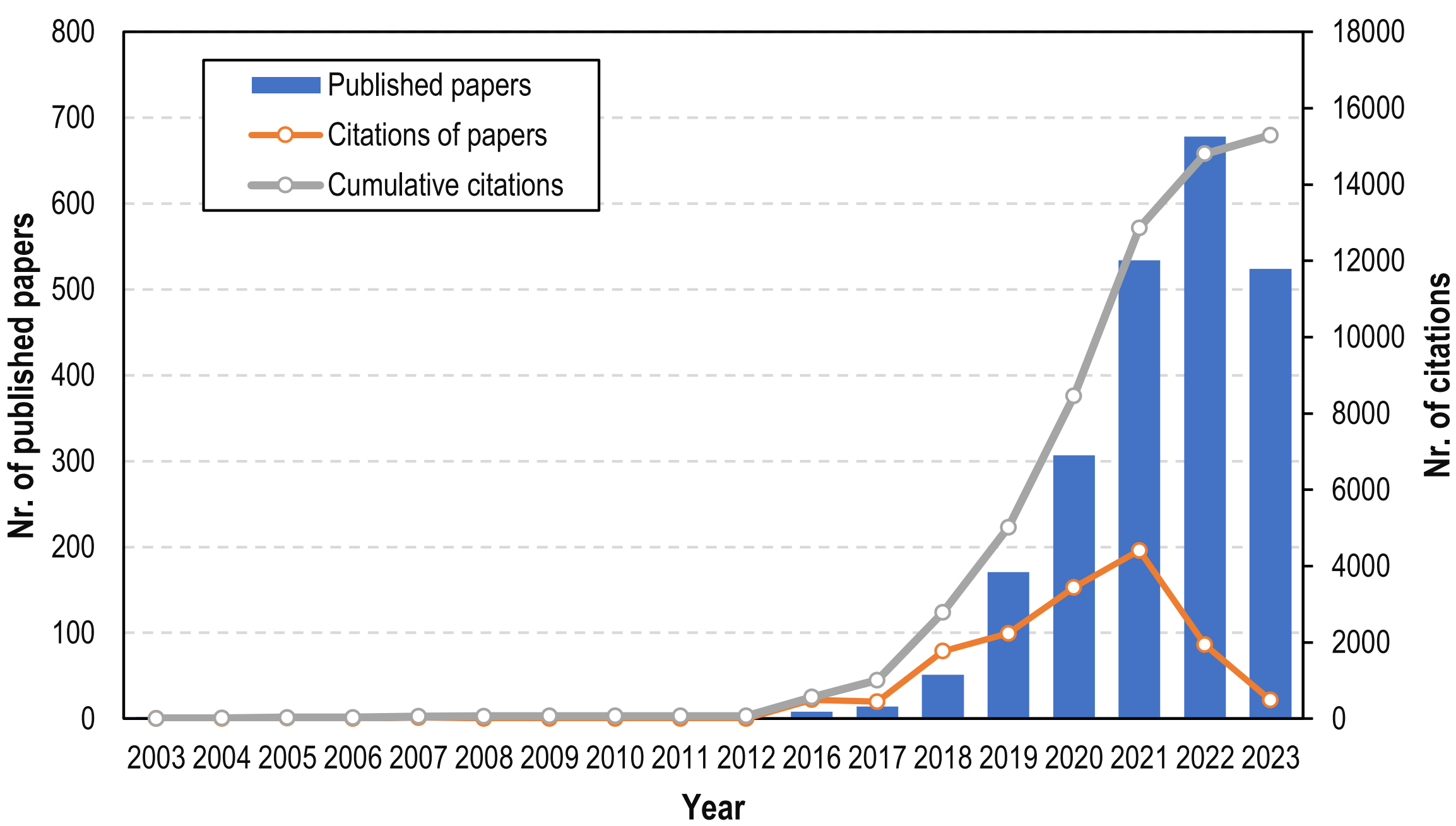

Figure 3 Research interest (number of papers) and research impact (number of citations) of RPA

The identification of linked subject areas was the focus of another analysis. Figure 4 shows the various subject areas where related publications were published. In the subject areas is first COMP (Computer Science) followed by BUSI (Business) and the top three is closed by ENGI (Engineering). COMP and ENGI both fall into the broader category of physical sciences. BUSI (Business) falls into the broader category of business management and accounting. Then, we can see DECI (Decision Sciences) and MATH (Mathematics), which we can associate with technical sciences. SOCI (Social Sciences) as a member of the humanistic category, we can observe multiple aspects of RPA, technical, business, and human (social) aspects of RPA. If we look at the perspective of citations, we can see that the business aspect and social aspects play a crucial role in robotic process automation. A lot of works cited precisely these two aspects of RPA. It supports the assumption that RPA is used as a tool for increasing the efficiency of business and helping companies with productivity. Not surprisingly the social aspect is also essential because the savings are many times made at the expense of the employee.

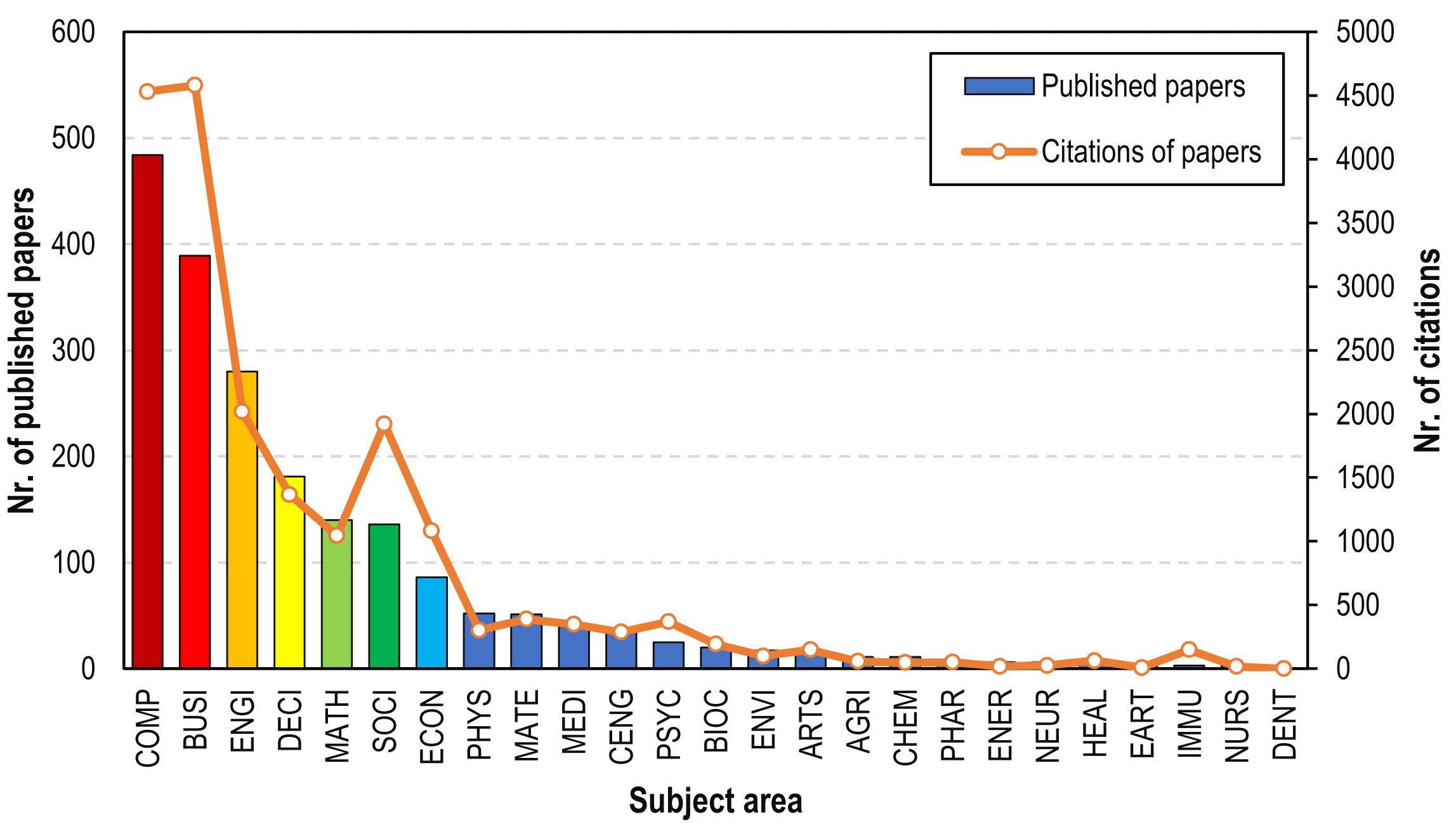

Figure 4 RPA related subject areas

## 3.2 Identified research topics

LDA topic modelling was used to answer RQ2. Based on the methodology described in section 2 we divided the corpus into 100 topics. According to several studies, it can be challenging to interpret so many different themes (Yeh et al., 2016). The outcome is illustrated in Figure 5.

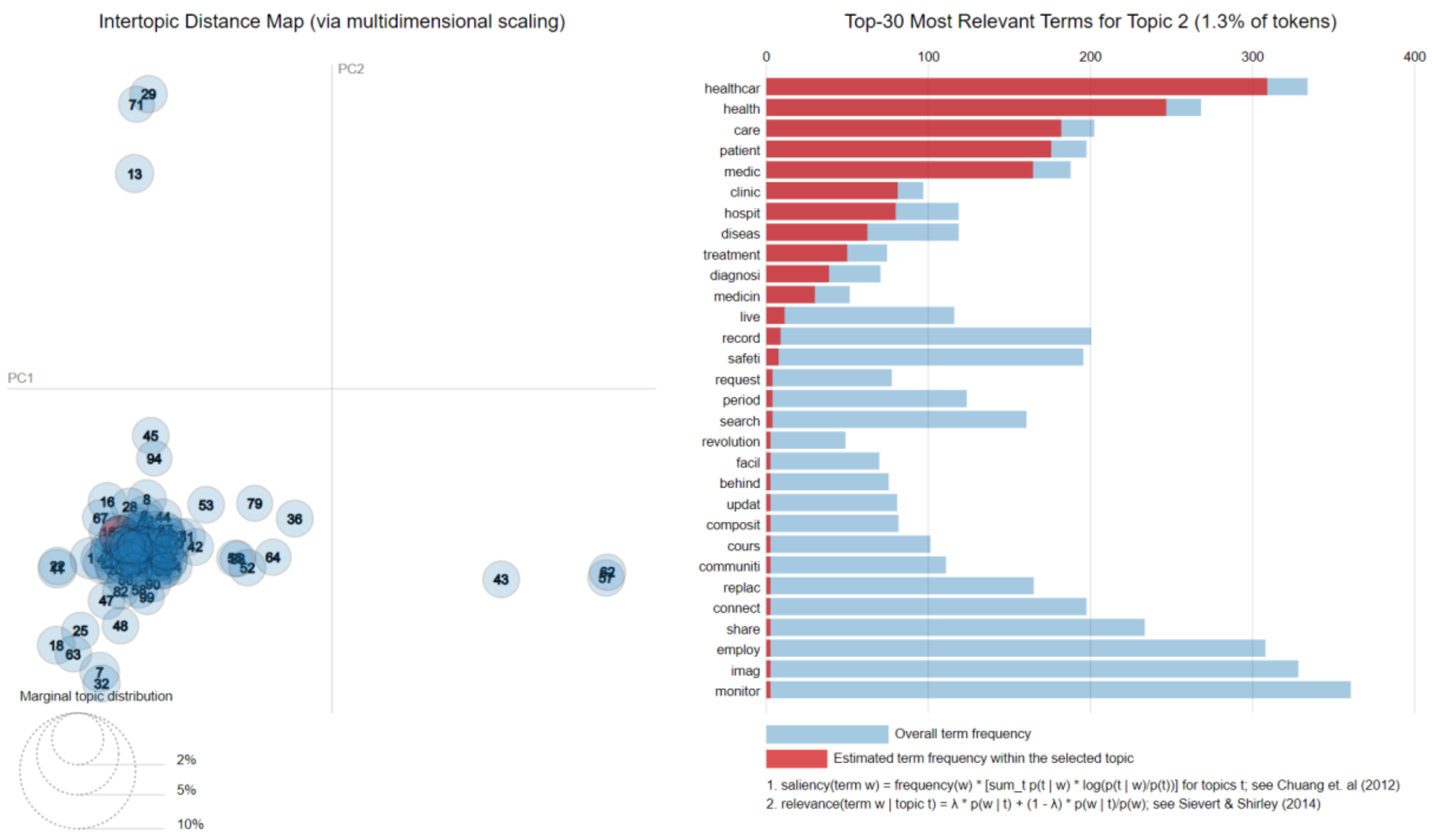

Figure 5 Intertopic distance map (right) and example of terms frequencies in a particular topic

A total of 100 separate topics were identified. De facto, there are 100 clusters, while each cluster (that is, topic) contains a certain number of papers. At the same time, each paper has a different recognition in the scientific community (measured by the number of citations). Figure 6 shows the overall topics description (upper part of Figure 4) and detailed topics description (lower part of Figure 6). Overall topics description contains four charts. The first of them is a combined donut chart (first from the left), which contains summary information on the number of citations in individual topics (the outer circle) and the number of articles in individual topics (the inner circle). From this chart, it can be seen that in some cases, the number of citations is disproportionately higher than the number of articles in the given topic - e.g. T-10, T-13 or T-32. However, this is the total number of citations, which can distort the results a bit, so we used to show CpP - (average) citations per paper. It is displayed together with the number of papers in the given topics on the radar chart (second from the left). It can be seen that the ratio between the number of papers and CpP is not equivalent, which means that some topics are more or less important in the field of RPA. A more detailed analysis of the most significant topics can be found in section 3.2. The third chart is a donut containing the top-10 subject areas in all topics (third from the left). The most numerous subject areas are COMP, BUSI, ENGI, and DECI. In the fourth bar chart (upper right), there is an overview of the top-30 most salient terms. These are the most numerous terms and their frequencies in the entire dataset. These terms had the greatest influence on the resulting composition of topics, which can be found in the lower part of Figure 6.

The characteristics of all 100 topics can be found at the bottom of Figure 6. This characteristic consists of topic number, number of papers, citation count, CpP, and top-5 terms in a particular topic. The right part contains bars showing the representation of individual subject areas in the given topic. Since some sources (e.g., journals, proceedings, etc.) were included in more than one subject area, the total sum of bars may be higher than the number of papers in a particular topic. However, the purpose of these bars is to point out the representation of individual subject areas in the given topics. For example, T-2 contains the terms healthcare; health; care; patient, and medic. Most papers in this topic are published in the subject area MEDI. The BUSI category is, for example, most represented in T-17, in which the most frequent terms are audit; auditor; firm; profess and client. The results of the comparison of the composition of top-5 terms and share of subject areas, therefore, have a logical justification.

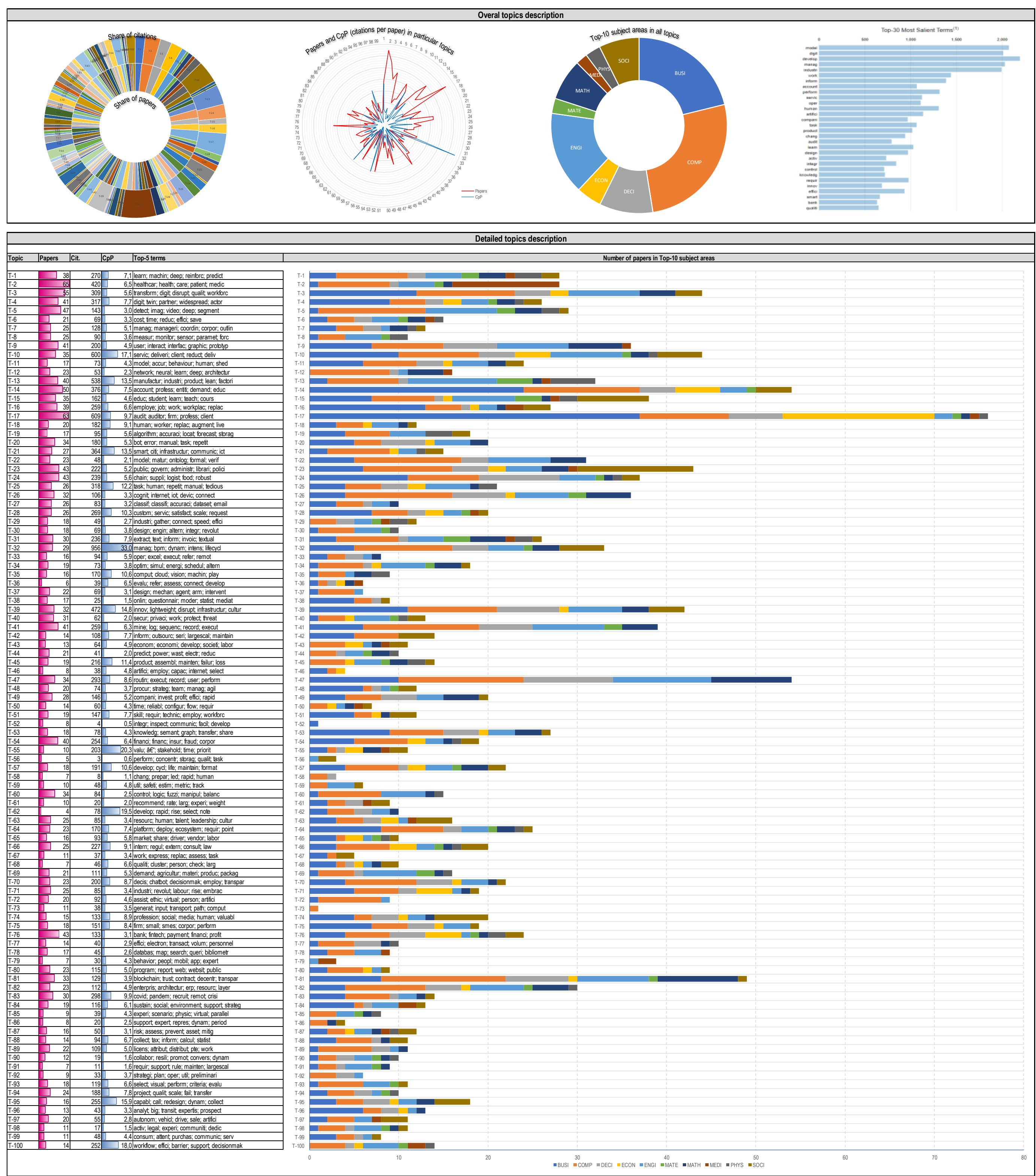

| Topic | Papers | Cit. | CpP | Top-5 terms |
|---|---|---|---|---|
| T-1 | 38 | 270 | 7,1 | learn; machin; deep; reinforc; predict |
| T-2 | 65 | 420 | 6,5 | healthcar; health; care; patient; medic |
| T-3 | 55 | 309 | 5,6 | transform; digit; disrupt; qualit; workforc |
| T-4 | 41 | 317 | 7,7 | digit; twin; partner; widespread; actor |
| T-5 | 47 | 143 | 3,0 | detect; imag; video; deep; segment |
| T-6 | 21 | 69 | 3,3 | cost; time; reduc; effici; save |
| T-7 | 25 | 128 | 5,1 | manag; manageri; coordin; corpor; outlin |
| T-8 | 25 | 90 | 3,6 | measur; monitor; sensor; paramet; forc |
| T-9 | 41 | 200 | 4,9 | user; interact; interfac; graphic; prototyp |
| T-10 | 35 | 600 | 17,1 | servic; deliveri; client; reduct; deliv |
| T-11 | 17 | 73 | 4,3 | model; accur; behaviour; human; shed |
| T-12 | 23 | 53 | 2,3 | network; neural; learn; deep; architectur |
| T-13 | 40 | 538 | 13,5 | manufactur; industri; product; lean; factori |
| T-14 | 50 | 376 | 7,5 | account; profess; entiti; demand; educ |
| T-15 | 35 | 162 | 4,6 | educ; student; learn; teach; cours |
| T-16 | 39 | 259 | 6,6 | employe; job; work; workplac; replac |
| T-17 | 63 | 609 | 9,7 | audit; auditor; firm; profess; client |
| T-18 | 20 | 182 | 9,1 | human; worker; replac; augment; live |
| T-19 | 17 | 95 | 5,6 | algorithm; accuraci; locat; forecast; storag |
| T-20 | 34 | 180 | 5,3 | bot; error; manual; task; repetit |
| T-21 | 27 | 364 | 13,5 | smart; citi; infrastructur; communic; ict |
| T-22 | 23 | 48 | 2,1 | model; matur; ontolog; formal; verif |
| T-23 | 43 | 222 | 5,2 | public; govern; administr; librari; polici |
| T-24 | 43 | 239 | 5,6 | chain; suppli; logist; food; robust |
| T-25 | 26 | 318 | 12,2 | task; human; repetit; manual; tedious |
| T-26 | 32 | 106 | 3,3 | cognit; internet; iot; devic; connect |
| T-27 | 26 | 83 | 3,2 | classif; classifi; accuraci; dataset; email |
| T-28 | 26 | 269 | 10,3 | custom; servic; satisfact; scale; request |
| T-29 | 18 | 49 | 2,7 | industri; gather; connect; speed; effici |
| T-30 | 18 | 69 | 3,8 | design; engin; altern; integr; revolut |
| T-31 | 30 | 236 | 7,9 | extract; text; inform; invoic; textual |
| T-32 | 29 | 956 | 33,0 | manag; bpm; dynam; intens; lifecycl |
| T-33 | 16 | 94 | 5,9 | oper; excel; execut; refer; remot |
| T-34 | 19 | 73 | 3,8 | optim; simul; energi; schedul; altern |
| T-35 | 16 | 170 | 10,6 | comput; cloud; vision; machin; play |
| T-36 | 6 | 39 | 6,5 | evalu; refer; assess; connect; develop |
| T-37 | 22 | 69 | 3,1 | design; mechan; agent; arm; intervent |
| T-38 | 17 | 25 | 1,5 | onlin; questionnair; moder; statist; mediat |
| T-39 | 32 | 472 | 14,8 | innov; lightweight; disrupt; infrastructur; cultur |
| T-40 | 31 | 62 | 2,0 | secur; privaci; work; protect; threat |
| T-41 | 41 | 259 | 6,3 | mine; log; sequenc; record; execut |
| T-42 | 14 | 108 | 7,7 | inform; outsourc; seri; largescal; maintain |
| T-43 | 13 | 64 | 4,9 | econom; economi; develop; societi; labor |
| T-44 | 21 | 41 | 2,0 | predict; power; wast; electr; reduc |
| T-45 | 19 | 216 | 11,4 | product; assembl; mainten; failur; loss |
| T-46 | 8 | 38 | 4,8 | artifici; employ; capac; internet; select |
| T-47 | 34 | 293 | 8,6 | routin; execut; record; user; perform |
| T-48 | 20 | 74 | 3,7 | procur; strateg; team; manag; agil |
| T-49 | 28 | 146 | 5,2 | compani; invest; profit; effici; rapid |
| T-50 | 14 | 60 | 4,3 | time; reliabl; configur; flow; requir |
| T-51 | 19 | 147 | 7,7 | skill; requir; technic; employ; workforc |
| T-52 | 8 | 4 | 0,5 | integr; inspect; communic; facil; develop |
| T-53 | 18 | 78 | 4,3 | knowledg; semant; graph; transfer; share |
| T-54 | 40 | 254 | 6,4 | financi; financ; insur; fraud; corpor |
| T-55 | 10 | 203 | 20,3 | valu; â€“; stakehold; time; priorit |
| T-56 | 5 | 3 | 0,6 | perform; concentr; storag; qualit; task |
| T-57 | 18 | 191 | 10,6 | develop; cycl; life; maintain; format |
| T-58 | 7 | 8 | 1,1 | chang; prepar; led; rapid; human |
| T-59 | 10 | 48 | 4,8 | util; safeti; estim; metric; track |
| T-60 | 34 | 84 | 2,5 | control; logic; fuzzi; manipul; balanc |
| T-61 | 10 | 20 | 2,0 | recommend; rate; larg; experi; weight |
| T-62 | 4 | 78 | 19,5 | develop; rapid; rise; select; note |
| T-63 | 25 | 85 | 3,4 | resourc; human; talent; leadership; cultur |
| T-64 | 23 | 170 | 7,4 | platform; deploy; ecosystem; requir; point |
| T-65 | 16 | 93 | 5,8 | market; share; driver; vendor; labor |
| T-66 | 25 | 227 | 9,1 | intern; regul; extern; consult; law |
| T-67 | 11 | 37 | 3,4 | work; express; replac; assess; task |
| T-68 | 7 | 46 | 6,6 | qualiti; cluster; person; check; larg |
| T-69 | 21 | 111 | 5,3 | demand; agricultur; materi; produc; packag |
| T-70 | 23 | 200 | 8,7 | decis; chatbot; decisionmak; employ; transpar |
| T-71 | 25 | 85 | 3,4 | industri; revolut; labour; rise; embrac |
| T-72 | 20 | 92 | 4,6 | assist; ethic; virtual; person; artifici |
| T-73 | 11 | 38 | 3,5 | generat; input; transport; path; comput |
| T-74 | 15 | 133 | 8,9 | profession; social; media; human; valuabl |
| T-75 | 18 | 151 | 8,4 | firm; small; smes; corpor; perform |
| T-76 | 43 | 133 | 3,1 | bank; fintech; payment; financi; profit |
| T-77 | 14 | 40 | 2,9 | effici; electron; transact; volum; personnel |
| T-78 | 17 | 45 | 2,6 | databas; map; search; queri; bibliometr |
| T-79 | 7 | 30 | 4,3 | behavior; peopl; mobil; app; expert |
| T-80 | 23 | 115 | 5,0 | program; report; web; websit; public |
| T-81 | 33 | 129 | 3,9 | blockchain; trust; contract; decentr; transpar |
| T-82 | 23 | 112 | 4,9 | enterpris; architectur; erp; resourc; layer |
| T-83 | 30 | 298 | 9,9 | covid; pandem; recruit; remot; crisi |
| T-84 | 19 | 116 | 6,1 | sustain; social; environment; support; strateg |
| T-85 | 9 | 39 | 4,3 | experi; scenario; physic; virtual; parallel |
| T-86 | 8 | 20 | 2,5 | support; expert; repres; dynam; period |
| T-87 | 16 | 50 | 3,1 | risk; assess; prevent; asset; mitig |
| T-88 | 14 | 94 | 6,7 | collect; tax; inform; calcul; statist |
| T-89 | 22 | 109 | 5,0 | licens; attribut; distribut; pte; work |
| T-90 | 12 | 19 | 1,6 | collabor; resili; promot; convers; dynam |
| T-91 | 7 | 11 | 1,6 | requir; support; rule; mainten; largescal |
| T-92 | 9 | 33 | 3,7 | strategi; plan; oper; util; preliminari |
| T-93 | 18 | 119 | 6,6 | select; visual; perform; criteria; evalu |
| T-94 | 24 | 188 | 7,8 | project; qualit; scale; fail; transfer |
| T-95 | 16 | 255 | 15,9 | capabl; call; redesign; dynam; collect |
| T-96 | 13 | 43 | 3,3 | analyt; big; transit; expertis; prospect |
| T-97 | 20 | 55 | 2,8 | autonom; vehicl; drive; sale; artifici |
| T-98 | 11 | 17 | 1,5 | activ; legal; experi; communiti; dedic |
| T-99 | 11 | 48 | 4,4 | consum; attent; purchas; communic; serv |
| T-100 | 14 | 252 | 18,0 | workflow; effici; barrier; support; decisionmak |

Figure 6 Research topics of RPA

## 3.3 The most significant topics and their evolution over time

To investigate deeper the RPA topic's influence - and for RQ3 answering - we plotted the subjects on a scatter diagram. According to research interest and research impact, Figure 6 displays all 100 topics. The size of the bubble represents the percentage growth in the number of articles over the past four years from 2019. The figure is divided into three levels (vertically and horizontally) by the addition of boundaries, representing the median (blue lines), third quartile (green lines), and 90th percentile (yellow lines). The subjects in each part are identified by a different colour to help in the analysis. Three issues (topics no. 1, 3, and 8) differentiated out from the rest for having either a much higher research impact (topic no. 8), a significantly higher research interest (topic no. 1), or both (topic no. 3).

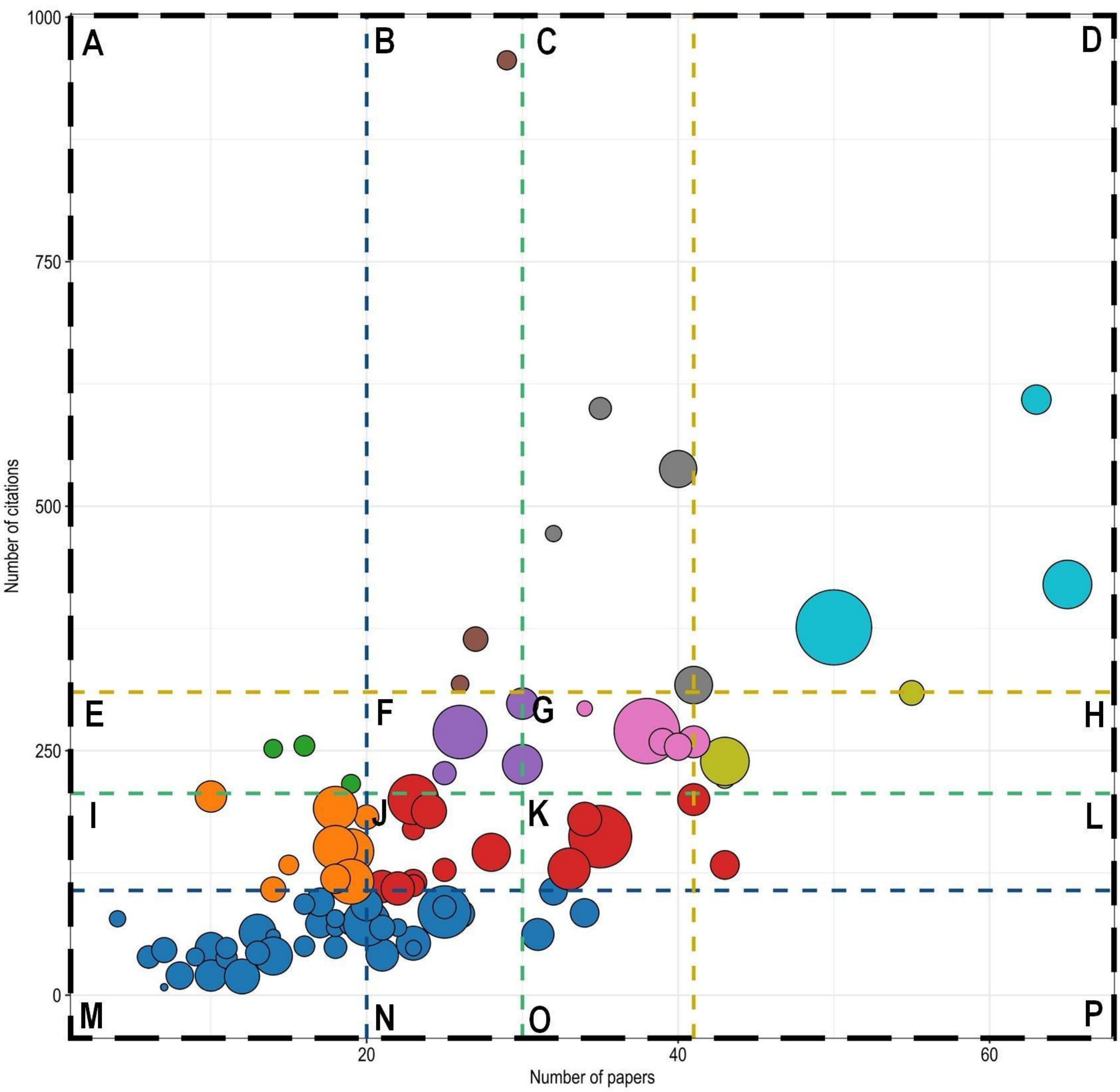

Figure 7 Scatter plot of research topics based on the number of their research interest (nr. of papers) and research impact (nr. of citations)

The divide, as noted above, was created so that we could more accurately assess the significance of specific themes because there are significant differences between the particular topics in terms of the relevance of research interest and research impact. We will discuss the most significant subjects (section D in Figure 7), themes with the most research interest (sections D, H, L, and P), and topics with the greatest research effect (sections A, B, C, and D) in the sections that follow.

The most significant themes were those with more papers and more citations than the third quartile (parts C, D, G, and H in Figure 7) within the same period. By focusing only on topics above the third quartile, we reveal the most significant topics related to RPA. From this, it can be inferred that these are subjects with a high level of research interest and research effect in the area of robotic process automation. These themes are included in Table 2.

| Topic nr. | Topic Label | Top-5 words | Research Interest (nr. of papers) | Research Impact (nr. of citations) |
| --- | --- | --- | --- | --- |
| 2 | Health Care | healthcar','health','care','patient','medic | 65 | 420 |
| 3 | Digital Transformation | transform','digit','disrupt','qualit','workforc | 55 | 309 |
| 4 | Digital Twin | digit','twin','partner','widespread','actor | 41 | 317 |
| 5 | Computer Vision | detect','imag','video','deep','segment | 47 | 143 |
| 10 | Service Delivery | servic','deliveri','client','reduct','deliv | 35 | 600 |
| 13 | Manufacturing | manufactur','industri','product','lean','factori | 40 | 538 |
| 14 | Accounting | account','profess','entiti','demand','educ | 50 | 376 |
| 17 | Audit | audit','auditor','firm','profess','client | 63 | 609 |
| 21 | Smart City | smart','citi','infrastructur','communic','ict | 27 | 364 |
| 23 | Public Sector | public','govern','administr','librari','polici | 43 | 222 |
| 24 | Supply Chain | chain','suppli','logist','food','robust | 43 | 239 |
| 25 | Task Analysis | task','human','repetit','manual','tedious | 26 | 318 |
| 32 | Life Cycle Assessment | manag','bpm','dynam','intens','lifecycl | 29 | 956 |
| 39 | Disruptive Technology | innov','lightweight','disrupt','infrastructur','cultu r | 32 | 472 |
| 76 | Banking | bank','fintech','payment','financi','profit | 43 | 133 |

***Table 2.*** *The most significant topics in robotic process automation*

The table contains the 15 most significant topics related to the robotic process automation area. The dominant position from the point of view of research impact is N32 life cycle assessment, which is by far the most cited topic. The reason may be that it is still a new technology and it is crucial for a feasibility study in the company, deployment to production, for keeping running in production, and evaluation of the benefits. The second and third topics have almost similar numbers of citations, topics of audit and service delivery.

The highest research interest is health care (N2). There is the most publication in the healthcare industry because there is huge potential for digitalization and improving processes. Also audit is really important, popular with 63 publications which is two publications less than the first health care. The third topic by research interest is digital transformations, with 55 publications.

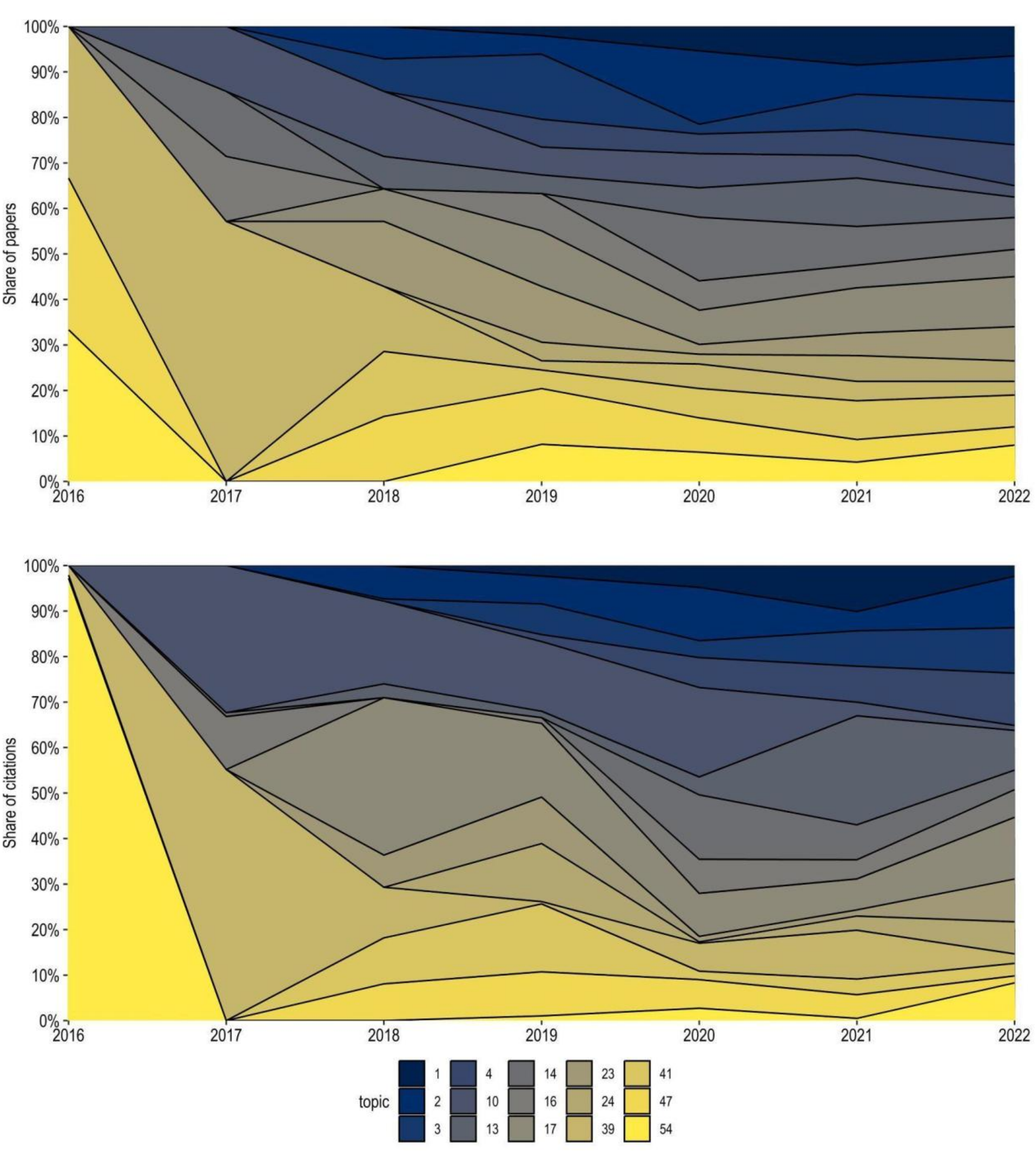

Figure 8. Evolution of the most significant topics (i. e., number of citations and number of papers was above the 3rd quartile)

From Figure 8 we can see that topic 39 (Disruptive technology) in the early days of 2017 technology, there were many publications about RPA being a distributive technology and it had one of the most significant weights. Topic 24, about the use of RPA in the Supply Chain started to emerge more slowly with a delay in 2019. Topic 1 Machine Learning and Artificial Intelligence only started to emerge in 2018 when it then gained importance in RPA. Topic 2 RPA in health care started to emerge around 2017 and steadily gained importance in the RPA field now is comparable with other topics. Topic 13 (Manufacturing) experienced a boom in the year after 2020 when it started to grow very much in citations. Before that, it was not as interesting and not many publications were generated. Topic 10 Service delivery is almost disappearing in the number of citations and its citation share is decreasing very much in relation to the others. Topic 17 Audit was very popular in RPA and then its share was decreasing relative to the others, but it's still an interesting topic for RPA. Topic 39 is disruptive technology by citation rate, it looks like it is not so disruptive anymore and fewer people are citing it, which shows that RPA is becoming more mainstream.

Assignment of individual papers to specific topics took place on the basis of calculated probability. According to LDA, each paper was assigned to each of the 100 topics with a certain probability. By comparing these probabilities, it was possible to identify the degree of independence of the topics. This was realised through bivariate correlation analysis, while the results of the Pearson correlation coefficient are found in the chord diagram in Figure 7 (Krzywinski et al., 2009) - only the relationships between the 15 most significant topics are shown. Overall, only very small values of correlation coefficients were identified. The highest correlation coefficients were identified between T-03 and T-04 (r=0.150), between T-10 and T-23 (r=0.117) and between T-14 and T-17 (r=0.101). This is a very low-intensity relationship, which means that the topics are well distinguishable from each other, and the LDA approach was able to identify unique topics.

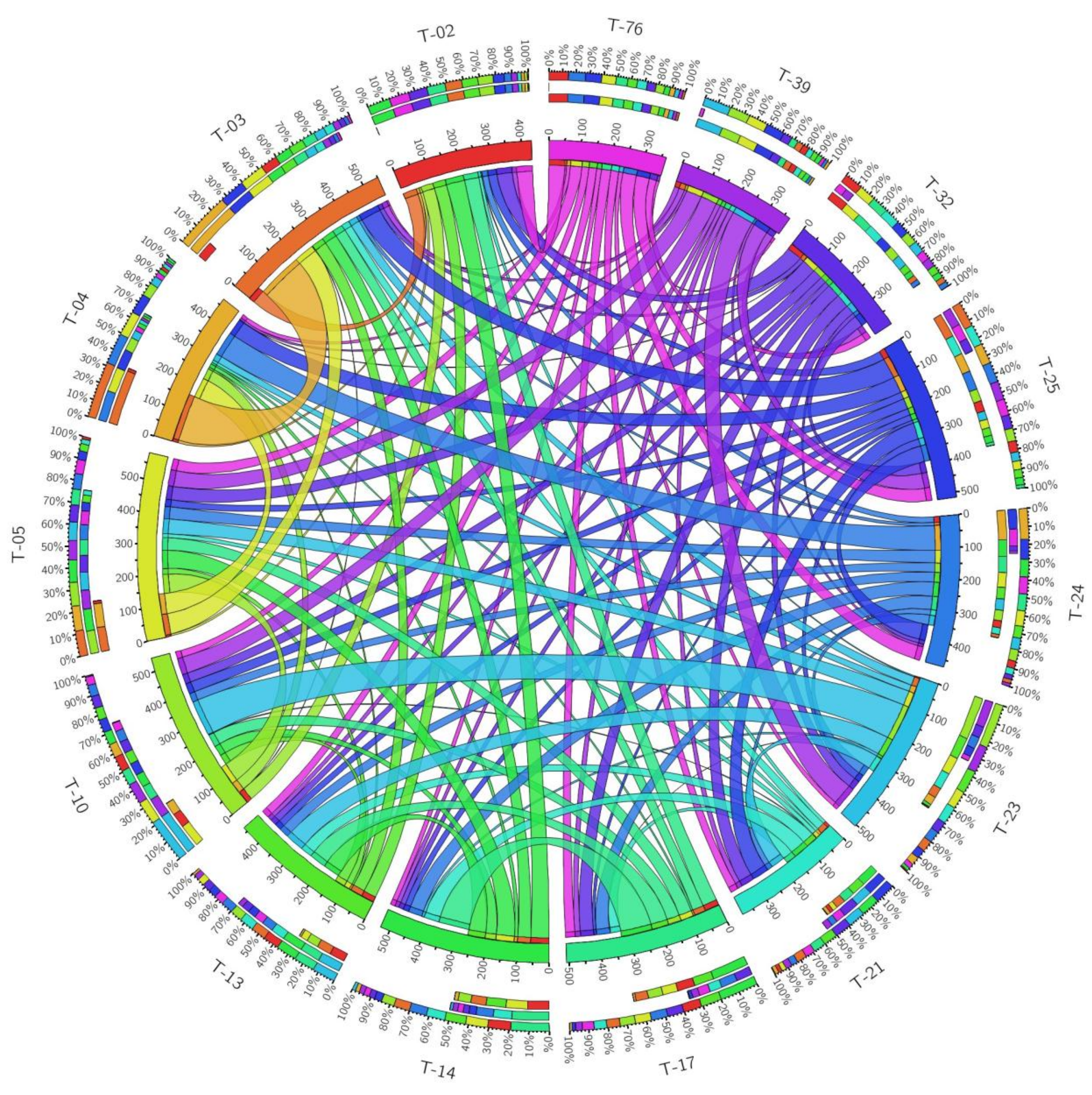

Figure 9 Correlation among topics of RPA

# 4 Discussion

Combining an in-depth exploration of contemporary trends with scrutiny of academic literature, our study offers practical insights and identifies research opportunities, striking a balance between immediate applicability and theoretical relevance within the context of the analysed topic.

## 4.1 Theoretical implications

The scientific community's growing interest in robotic process automation (RPA) is evidence of the technology's disruptive potential. The expanding importance of RPA is particularly highlighted by the increasing number of publications and citations. It's clear that RPA is becoming recognized as a force for revolutionary change as its user base keeps growing in tandem with the number of businesses implementing this technology (Haleem et al., 2021). This explosion is shown not only in Figure 3, which shows the market and RPA publications growing at the same time, but also enables the creation of a new industry of hyper-automation.

From the analysis of the most relevant topics it can be clearly seen that RPA serves as an agent for the automation of processes or as a driver of digital transformation/change. With most use cases RPA was used together with artificial intelligence, and before deployment is many time used tools and technologies from the business process management field. From the analysis follows that RPA is deployed with artificial intelligence and has become a crucial complementary technology. The newly emerged hyper-automation trend which connects all automation tools together and helps companies increase their efficiency and effectiveness. The field of hyper-automation has new potential for future research. There is a clear potential development of cooperation in artificial intelligence with robotic process automation, which can serve as an agent for the deployment of AI in organisations.
Research potential in the field of hyper-automation:

1) Map the hyper-automation journey, tools, technologies, approaches and use cases.
2) Finding similar attributes among industries that apply and use RPA.
3) Make digital transformation easier in regulated industries such as health care or the public sector.

The most significant topics that were identified in our research were examined in previous studies. Syed et al. (2020) briefly discuss that computer vision (topic nr. 5) is a complementary technology to RPA. Next discuss topics of digital transformation (topic nr. 3), service delivery (topic nr. 10), life cycle assessment benefits (topic nr. 32), relation to the RPA and links between each other. Also, propose a research challenge on the topic of task analysis (topic nr. 25), that is crucial to access better task selection.

Enriquez et al. (2020), except for discussion of life cycle assessment and computer vision, presented that over 7% of articles are related to the use of RPA in the industry as health care (topic nr. 2), public sector (topic nr. 23) and others. Life cycle assessment and finances (finances is a connection of banking, accounting and auditing - topics nr. 76, 14 and 17) make around 43% of articles.

Stravinskienė & Serafinas (2021) focus on grouping RPA papers if they are from the field of manufacturing (topic nr. 13) or service delivery (topic nr. 10). Resulting that 41% is from a field of services, approximately 7% from manufacturing and 28% are categorised as from both. Also mentioned is that the application of RPA in accounting (topic nr. 14), supply chain (topic nr. 24), and finance is covered in studies (Anagnoste, 2017a, 2017b; Hartley & Sawaya, 2019).

Wewerka & Reichert, (2021), in a systematic review, discuss the importance of Life Cycle Assessment (topic nr. 32) and applying RPA in the banking sector and using RPA in audit (topic nr. 17). Mention the implementation of RPA in manufacturing fields where processes are digitalized.

Digital twin (topic nr. 4) and Smart City (topic nr. 21) topics have not been thoroughly explored and remain uncharted in RPA research, lacking systematic study.

## 4.2 Practical implications

We have identified practical implications and potential research opportunities emerging from the analysis.

The first practical implication is the deployment of RPA within artificial intelligence. We can expect the growth of more publications on this topic. The trend towards generative AI, led by Large Language Models (LLMs), has attracted a lot of attention from the public and large companies. Already, many fields, both commercial and non-commercial, had to respond to this trend of new generative AI. The combination of RPA and LLMs will make it much easier to work with RPA when AI can help to decode RPA code, automate new processes, and create new intelligent assistants. For the past few years, some researchers have been trying to create intelligent assistants by combining RPA with AI (Rizk et al., 2020). However, now with LLMs, some things will be much easier to do, and AI is an enabler of new possibilities in this regard (Huo et al., 2023). Outside the regards of LLMs there is still room for decision-making in processes led by AI. Where RPA needs precise information to make a decision with AI the decision can be closer to human decision. The future of orchestration hyper-automation and all automated processes with RPA or without are hard to operate manually to maximise the potential of the automation.
Research potential with AI topic:

1) Application LLMs in the RPA domain (LLM models facilitate and render more feasible the execution of this type of research, though it is not solely reliant on LLMs; it could be accomplished without them, albeit with greater difficulty.):
    a) for automatic generation of the code (user text conversion to code),
    b) improving and refactoring the RPAs code, exception handling, and automatic repairs,
    c) improvement understandability of RPA for citizen development and non-technical users
2) Optimization of operations of hyper-automation and RPA hence orchestration. To maximise value from licences and automation infrastructure. Ensure logical sequences without delays and errors in processes.
3) Improvement of decision-making of RPA robots to handle more stochastic processes.

In relation to stochastic processes RPA and AI is important to also mention the topic 41 task-mining. Before deployment of RPA is crucial to map the process and create a product design document (PDD) where task-mining is really handy. With this topic comes to another research potential with task-mining:

1) Improvement task-mining techniques to effectively handle stochastic tasks.
2) Automatic creation of PDD

On top of all these topics stands the topic of the social aspect of RPA. Considering the human aspect is a key element for the successful implementation of automation (Wewerka & Reichert, 2021). Current research shows interesting data that people are afraid of being replaced, but also increases people's happiness and motivation due to not performing boring tasks (Fernandez & Aman, 2018; Suri et al., 2019). Which do not create conflict. It brings another research potential of the topic of social aspects:

1) The immediate, medium-term, and long-term influence of Robotic Process Automation (RPA) on employees and other relevant stakeholders.
2) The impact of digital transformation on both employees and customers.

With the implication that RPA serves as an agent of digital transformation, there are some practical implications for easier digital transformation.

Research potential in the field of hyper-automation:

1) Finding applications and use cases in other administrative fields for RPA.
2) Enabling feasible citizen development in organisations and empowering employees to create automation.

## 4.3 Research limitations and future research directions

Every study has its weaknesses and this one is no different. The first restriction relates to how textual data is processed. The foundation for topic modelling was the documents that we found by employing certain keywords. This modelling processed document abstracts and not complete full texts, which can be a restriction. The abstract contains all the pertinent data necessary to correspond to the content of the specific paper, and occasionally the abstract may be overly wide or imprecise, which undermines the legitimacy of the chosen subjects.

Another limitation is related to our aim to examine every paper about robotic process automation published from the start of this field of study in the Scopus and Web of Science databases. We entered the appropriate search term as defined by Verhoef et al. (2021). Additionally, we included a definition of "robotic process automation" in the search term. Nevertheless, despite this, there is a slight chance that some papers from this field were left out of our analysis. However, we presume that since this is such a tiny and insignificant portion of documents, it has no significant impact on our findings.

The final drawback is the subjectivity involved in selecting stopwords during data pre-processing. The usage of words like hence, however, address, general, etc., that had no meaning and were useless for understanding the topics was prevalent in the abstracts. The subjective list of these stopwords may have influenced how the terms in particular themes were composed. Subjectivity is also related to the naming of revealed themes (section 3.3). Such naming was done using an analysis of the top 5 terms, but we do not rule out the possibility of changing the topic's name to something more appropriate after carefully examining linked publications.

These limitations and restrictions represent space for investigation and future research. Even after several years, the same methodology can be used to spot changes in topics, including their creation, absence, splitting, or merging. Our research focused on robotic process automation and research related to RPA. Our study shows multiple potential ways of applying RPA in various industries where further investigation is possible. Also future research could be extended to the use of RPA and artificial intelligence or investigation of the new field of hyper-automation.

# References

Anagnoste, S. (2017a). *Robotic Automation Process—The next major revolution in terms of back office operations improvement* (WOS:000431004400072). *11*(1), 676–686. https://doi.org/10.1515/picbe-2017-0072

Anagnoste, S. (2017b). Robotic Automation Process—The next major revolution in terms of back office operations improvement. *Proceedings of the International Conference on Business Excellence*, *11*(1), 676–686. https://doi.org/10.1515/picbe-2017-0072

Arun, R., Suresh, V., Veni Madhavan, C. E., & Narasimha Murthy, M. N. (2010). On Finding the Natural Number of Topics with Latent Dirichlet Allocation: Some Observations. In M. J. Zaki, J. X. Yu, B. Ravindran, & V. Pudi (Eds.), *Advances in Knowledge Discovery and Data Mining* (Vol. 6118, pp. 391–402). Springer Berlin Heidelberg. https://doi.org/10.1007/978-3-642-13657-3_43

Asmussen, C. B., & Møller, C. (2019). Smart literature review: A practical topic modelling approach to exploratory literature review. *Journal of Big Data*, *6*(1), 93. https://doi.org/10.1186/s40537-019-0255-7

Blei, D. M., & Lafferty, J. D. (2007). A correlated topic model of Science. *The Annals of Applied Statistics*, *1*(1), 17–35. https://doi.org/10.1214/07-AOAS114

Blei, D. M., & Lafferty, J. D. (2009). Topic models. In *Text mining* (pp. 101–124). Chapman and Hall/CRC.

Blei, D. M., Ng, A. Y., & Jordan, M. I. (2003). Latent Dirichlet Allocation. *J. Mach. Learn. Res.*, *3*(null), 993–1022.

Cao, J., Xia, T., Li, J., Zhang, Y., & Tang, S. (2009). A density-based method for adaptive LDA model selection. *Neurocomputing*, *72*(7–9), 1775–1781. https://doi.org/10.1016/j.neucom.2008.06.011

Czarnecki, C., & Fettke, P. (Eds.). (2021). *Robotic process automation: Management, technology, applications*. De Gruyter Oldenbourg.

Deveaud, R., SanJuan, E., & Bellot, P. (2014). Accurate and effective latent concept modeling for ad hoc information retrieval. *Document Numérique*, *17*(1), 61–84. https://doi.org/10.3166/dn.17.1.61-84

*Download RPA Listings—RPA Snippets, Workflows, Connectors*. (n.d.). UiPath Marketplace. Retrieved August 15, 2023, from https://marketplace.uipath.com/listings?sort=newest

Enriquez, J. G., Jimenez-Ramirez, A., Dominguez-Mayo, F. J., & Garcia-Garcia, J. A. (2020). Robotic Process Automation: A Scientific and Industrial Systematic Mapping Study. *IEEE Access*, *8*, 39113–39129. https://doi.org/10.1109/ACCESS.2020.2974934

Fernandez, D., & Aman, A. (2018). Impacts of Robotic Process Automation on Global Accounting Services. *Asian Journal of Accounting and Governance*, *9*, 123–132. https://doi.org/10.17576/AJAG-2018-09-11

Fersht, P. (2018, November 30). RPA will reach $2.3bn next year and $4.3bn by 2022... As we revise our forecast upwards. *Horses for Sources | No Boundaries*. https://www.horsesforsources.com/rpa-forecast-2016-2022_120118/

Gartner Inc. (2021, December 16). *Move beyond RPA to deliver hyperautomation*. Gartner. https://www.gartner.com/en/doc/433853-move-beyond-rpa-to-deliver-hyperautomation

Griffiths, T. L., & Steyvers, M. (2004). Finding scientific topics. *Proceedings of the National Academy of Sciences*, *101*(suppl_1), 5228–5235. https://doi.org/10.1073/pnas.0307752101

Grün, B., & Hornik, K. (2011). topicmodels: An R Package for Fitting Topic Models. *Journal of Statistical Software*, *40*(13). https://doi.org/10.18637/jss.v040.i13

Haleem, A., Javaid, M., Singh, R. P., Rab, S., & Suman, R. (2021). Hyperautomation for the enhancement of automation in industries. *Sensors International*, *2*, 100124. https://doi.org/10.1016/j.sintl.2021.100124

Hartley, J. L., & Sawaya, W. J. (2019). Tortoise, not the hare: Digital transformation of supply chain business processes. *Business Horizons*, *62*(6), 707–715. https://doi.org/10.1016/j.bushor.2019.07.006

Hewitt, L. R., Richard-Noel, T., & Shukla, N. M. (2021). *A vision of how Robotic Process Automation (RPA) can be a force for good across the global healthcare industry, its impacts, benefits, and opportunities. This paper offers a detailed perspective from the UK's National Health Service (NHS) and their RPA Centres of Excellence, and describes the value of RPA for connectivity, collaboration, and productivity across healthcare.*

Holmberg, M., & Dobslaw, F. (2022). An Industrial Case-Study on GUI Testing With RPA. *2022 Ieee 15th International Conference on Software Testing, Verification and Validation Workshops (Icstw 2022)*, 199–206. https://doi.org/10.1109/ICSTW55395.2022.00043

Huo, S., Mukherjee, K., Bandlamudi, J., Isahagian, V., Muthusamy, V., & Rizk, Y. (2023). Accelerating the Support of Conversational Interfaces for RPAs Through APIs. In J. Köpke, O. López-Pintado, R. Plattfaut, J.-R. Rehse, K. Gdowska, F. Gonzalez-Lopez, J. Munoz-Gama, K. Smit, & J. M. E. M. Van Der Werf (Eds.), *Business Process Management: Blockchain, Robotic Process Automation and Educators Forum* (Vol. 491, pp. 165–180). Springer Nature Switzerland. https://doi.org/10.1007/978-3-031-43433-4_11

Ivančić, L., Suša Vugec, D., & Bosilj Vukšić, V. (2019). Robotic Process Automation: Systematic Literature Review. In C. Di Ciccio, R. Gabryelczyk, L. García-Bañuelos, T. Hernaus, R. Hull, M. Indihar Štemberger, A. Kő, & M. Staples (Eds.), *Business Process Management: Blockchain and Central and Eastern Europe Forum* (Vol. 361, pp. 280–295). Springer International Publishing. https://doi.org/10.1007/978-3-030-30429-4_19

Krzywinski, M., Schein, J., Birol, İ., Connors, J., Gascoyne, R., Horsman, D., Jones, S. J., & Marra, M. A. (2009). Circos: An information aesthetic for comparative genomics.

*Genome Research*, *19*(9), 1639–1645. https://doi.org/10.1101/gr.092759.109

Lacity, M., & Willcocks, L. (2016). Robotic Process Automation at Telefonica O2. *MIS QUARTERLY EXECUTIVE*, *15*(1), 21–35.

Lacity, M., Willcocks, L., & Craig, A. (2016). Robotizing Global Financial Shared Services at Royal DSM. *The Outsourcing Unit Working Research Paper Series*, 26.

Lhuer, X. (2016). *The next acronym you need to know about: RPA (robotic process automation)*. https://www.mckinsey.com/capabilities/mckinsey-digital/our-insights/the-next-acronym-you-need-to-know-about-rpa

Min, C., Sun, J., Pei, L., & Ding, Y. (2016). *Measuring delayed recognition for papers: Uneven weighted summation and total citations*. https://doi.org/10.1016/j.joi.2016.10.001

Ponweiser, M. (2012). *Latent Dirichlet Allocation in R*. https://api.semanticscholar.org/CorpusID:60668264

Rizk, Y., Isahagian, V., Boag, S., Khazaeni, Y., Unuvar, M., Muthusamy, V., & Khalaf, R. (2020). A Conversational Digital Assistant for Intelligent Process Automation. In A. Asatiani, J. M. García, N. Helander, A. Jiménez-Ramírez, A. Koschmider, J. Mendling, G. Meroni, & H. A. Reijers (Eds.), *Business Process Management: Blockchain and Robotic Process Automation Forum* (Vol. 393, pp. 85–100). Springer International Publishing. https://doi.org/10.1007/978-3-030-58779-6_6

*Robotic Process Automation Market Size & Share Report 2030.* (n.d.). Spherical Insights. Retrieved August 15, 2023, from https://www.sphericalinsights.com/reports/robotic-process-automation-market

Stravinskienė, I., & Serafinas, D. (2021). Process Management and Robotic Process Automation: The Insights from Systematic Literature Review. *Management of Organizations: Systematic Research*, *85*(1), 87–106. https://doi.org/10.1515/mosr-2021-0006

Suri, V. K., Elia, M. D., Arora, P., & Van Hillegersberg, J. (2019). Automation of Knowledge-Based Shared Services and Centers of Expertise. In J. Kotlarsky, I. Oshri, & L.

Willcocks (Eds.), *Digital Services and Platforms. Considerations for Sourcing* (Vol. 344, pp. 56–75). Springer International Publishing. https://doi.org/10.1007/978-3-030-15850-7_4

Syed, R., Suriadi, S., Adams, M., Bandara, W., Leemans, S. J. J., Ouyang, C., ter Hofstede, A. H. M., van de Weerd, I., Wynn, M. T., & Reijers, H. A. (2020). Robotic Process Automation: Contemporary themes and challenges. *Computers in Industry*, *115*, 103162. https://doi.org/10.1016/j.compind.2019.103162

Vailshery, L. S. (2022, June 1). *Global RPA market size 2030*. Statista. https://www.statista.com/statistics/1309384/worldwide-rpa-software-market-size/

van der Aalst, W. M. P., Bichler, M., & Heinzl, A. (2018). Robotic Process Automation. *Business & Information Systems Engineering*, *60*(4), 269–272. https://doi.org/10.1007/s12599-018-0542-4

Virbahu Nandishwar Jain & DAZ Systems. (2019). Robotics for Supply Chain and Manufacturing Industries and Future It Holds! *International Journal of Engineering Research And*, *V8*(03), IJERTV8IS030062. https://doi.org/10.17577/IJERTV8IS030062

Wewerka, J., & Reichert, M. (2021). Robotic process automation—A systematic mapping study and classification framework. *Enterprise Information Systems*, 1–38. https://doi.org/10.1080/17517575.2021.1986862

Willcocks, L. P., Lacity, M., & Craig, A. (2015). *The IT function and robotic process automation* (Issue 64519, pp. 1–39). London School of Economics and Political Science, LSE Library. https://ideas.repec.org/p/ehl/lserod/64519.html

# Appendix A

"process", "autom", "technolog", "use", "rpa", "robot", "data", "research", "system", "studi", "busi", "paper", "intellig", "applic", "result", "provid", "implement", "base", "approach", "author", "propos", "analysi", "method", "organ", "improv", "increas", "present", "challeng", "identifi", "case", "effect", "differ", "softwar", "tool", "adopt", "impact", "literatur", "find", "natur", "practic", "futur", "includ", "framework", "solut", "review", "show", "current", "howev", "relat", "discuss", "signific", "aim", "import", "techniqu", "focus", "reserv", "right", "various", "articl", "potenti", "mani", "understand", "test", "sector", "springer", "exist", "field", "emerg", "ieee", "appli", "level", "benefit", "main", "structur", "factor", "advanc", "problem", "limit", "analyz", "role", "environ", "consid", "function", "area", "complex", "explor", "contribut", "purpos", "success", "specif", "publish", "issu", "conduct", "switzerland", "enabl", "object", "creat", "achiev", "investig", "high", "recent", "context", "order", "concept", "document", "examin", "posit", "within", "addit", "sever", "associ", "key", "address", "allow", "year", "becom", "influenc", "part", "due", "three", "construct", "type", "need", "possibl", "event", "opportun", "build", "therefor", "scienc", "combin", "exclus", "compar", "topic", "offer", "demonstr", "continu", "organiz", "term", "methodolog", "indic", "number", "relationship", "enhanc", "direct", "perspect", "better", "survey", "introduc", "automat", "train", "make", "chapter", "trend", "gap", "domain", "relev", "among", "respons", "lead", "toward", "implic", "insight", "action", "regard", "follow", "global", "world", "general", "institut", "initi", "thus", "interview", "suggest", "languag", "featur", "systemat", "involv", "expect", "adapt", "tradit", "aspect", "group", "describ", "question", "analys", "competit", "avail", "affect", "empir", "theori", "obtain", "goal", "form", "final", "conceptu", "highlight", "academ", "ltd", "major", "determin", "face", "accord", "critic", "individu", "countri", "still", "standard", "theoret", "interest", "valid", "advantag", "state", "exampl", "concern", "carri", "complet", "perceiv", "compon", "multipl", "especi", "common", "thing", "sourc", "recognit", "organis", "must", "novel", "across", "practition", "access", "sampl", "stage", "help", "higher", "without", "effort", "defin", "procedur", "book", "comprehens", "suitabl", "characterist", "condit", "furthermor", "contain", "mean", "univers", "previous", "essenti", "dimens", "reveal", "outcom", "consist", "best", "grow", "wide", "detail", "given", "overal", "real", "four", "flexibl", "view", "ensur", "second", "overview", "growth", "sinc", "gain", "open", "abl", "take", "set", "modern", "establish", "facilit", "leverag", "scientif", "usag", "engag", "solv", "â€“", "explain", "rang", "promis", "caus", "descript", "made", "start", "element", "compet", "accept", "step", "consequ", "consider", "abil", "depend", "variabl", "principl", "extend", "bring", "observ", "subject", "situat", "necessari", "originalityvalu", "way", "pattern", "special", "copyright", "deriv", "identif", "hand", "line", "handl", "now", "designmethodologyapproach", "via", "whether", "meet", "low", "along", "done", "realiz", "content", "phase", "link", "basi", "evid", "elsevi", "illustr", "unit", "moreov", "capit", "decad", "total", "idea", "evolv", "journal", "yet", "guid", "name", "respond", "introduct", "amount", "era", "earli", "creation", "although", "emerald", "deal", "conclud", "particular", "cover", "reason", "remain", "similar", "percept", "give", "realworld", "respect", "space", "appropri", "precis", "five", "primari", "place", "today", "minim", "center", "discov", "acceler", "less", "origin", "equip", "target", "confirm", "realiti", "singapor", "size", "captur", "intent", "incorpor", "nation", "crucial", "overcom", "paradigm", "substanti", "core", "progress", "evolut", "certain", "quick", "scope", "fundament", "reflect", "proceed", "offic", "despit", "central", "simpl", "basic", "known", "numer", "degre", "correct", "cooper", "around", "readi", "uncertainti", "depart", "discoveri", "confer", "align", "american", "rulebas", "shown", "popular", "alreadi", "past", "henc", "scholar", "everi", "output", "definit", "greater", "next", "receiv", "entir", "conclus", "seek", "difficult", "taken", "background", "indepth", "theme", "strong", "divers", "creativ", "lower", "run", "intend", "faster", "long", "typic", "interpret", "disciplin", "awar", "close", "list", "mdpi", "stateoftheart", "uniqu", "reli", "basel", "manner", "avoid", "inc", "aibas", "full", "per", "answer", "daili", "expand", "six", "other", "actual", "singl", "keep", "togeth", "desir", "decreas", "clear", "reach", "averag", "prove", "varieti", "anoth", "nowaday", "guidelin", "driven", "status", "constant", "proper", "top", "beyond", "european", "rais", "littl", "extent", "either", "partial", "fill", "emphas", "least", "short", "index", "assign", "fourth", "know", "section", "fact", "adjust", "appear", "like", "properti", "prefer", "come", "subsequ", "china", "built", "under", "class", "return", "gmbh", "acquir", "vari", "mix", "attract", "attempt", "upon", "journey", "whole", "just", "retriev", "alway", "taylor", "represent", "summar", "press", "scalabl", "difficulti", "elabor", "inclus", "character", "brought", "phenomenon", "easili", "lack", "franci", "enter", "occur", "almost", "look", "add", "hold", "huge", "india", "worldwid", "usual", "begin", "llc", "limitationsimpl", "specifi", "choic", "igi", "regular", "longterm", "charact", "one", "throughout", "independ", "accomplish", "maxim", "usabl", "synthes", "claim", "miss", "meaning", "third", "thorough", "iter", "hard", "scopus", "seem", "holist", "ideal", "probabl", "besid", "most", "yield", "latest", "normal", "assur", "abstract", "acm", "interdisciplinari", "near", "opensourc", "googl", "wiley", "span", "maximum", "treat", "feder", "hypothes", "fulfil", "great", "conveni", "notion", "believ", "therebi", "todayâ€™", "john", "presenc", "coupl", "highest", "seven", "outperform", "centr", "adequ", "necess", "nevertheless", "tri", "enrich", "simplifi", "differenti", "think", "editor", "son", "contrast", "wherea", "academia", "inspir", "emphasi", "resolv", "insid", "walter", "seamless", "gruyter", "tremend", "centuri", "informa", "clarifi", "berlinboston"